\documentclass[aps,12pt]{revtex4}
\usepackage{amsmath,amssymb,graphicx,epsfig}
\begin{document}


\title{Implications of area scaling of quantum fluctuations}

\author{Ram Brustein$^{(1)}$, David H. Oaknin$^{(2)}$, Amos Yarom$^{(1)}$}

\address{(1) Department of Physics, Ben-Gurion University,
Beer-Sheva 84105, Israel \\
(2) Department of Physics and Astronomy, University of British
Columbia,
Vancouver V6T 1Z1, Canada \\
\centerline{ {\rm E-mail:} {\tt
ramyb@bgumail.bgu.ac.il,doaknin@physics.ubc.ca,
yarom@bgumail.bgu.ac.il }}}

\begin{abstract}
Quantum fluctuations of a certain class of bulk operators defined
in spatial sub-volumes of Minkowski space-time, have an unexpected
area scaling property. We wish to present evidence that such area
scaling may be ascribed to a boundary theory. We first highlight
the implications of area scaling with two examples in which the
boundary area of the spatial regions is not monotonous with their
volume. Next, we prove that the covariance of two operators that
are restricted to two different regions in Minkowski space scales
linearly with their mutual boundary area. Finally, we present an
example which demonstrates why this implies an underlying boundary
theory.
\end{abstract}

\maketitle

\section{Introduction}
Area scaling of black hole thermodynamics
\cite{BekensteinBHentropy,Hawking} has lead to a better
understanding of some of the properties that a quantum theory of
gravity should have, and to the concept of holography
\cite{holsusskind} or dimensional reduction \cite{holthooft}. One
formulation of the holographic principle states that $d+1$
dimensional quantum gravitational systems can be described by some
boundary theory living on a $d$ dimensional hyper-surface (see
\cite{Bousso} for a review). An explicit example of this is given
by the AdS/CFT correspondence \cite{AdSCFTrev}.

Considering not a curved, but flat Minkowski space-time,
one can show that when restricting measurements to a spatial
volume $V$, the variance of quantum fluctuations of (a certain class
of) bulk operators scales as the surface area of $V$
\cite{holmin}. These fluctuations are due to the entangled
properties of states inside and outside the region $V$.

Although these fluctuations are quantum mechanical in nature,
arising from expectation values of operators in a pure state $|0
\rangle$, for example $\langle 0 | O^V |0 \rangle$, an observer
restricted to the region $V$ will observe statistical fluctuations
that are determined by a density matrix $\rho_{in}$:
$\text{Tr}(\rho_{in}O_V)$ \cite{feynmanbookstatmech}. $\rho_{in}$
is obtained by taking the density matrix describing the original
state and tracing over the degrees of freedom external to $V$:
$\rho_{in}=\text{Tr}_{out}|0><0|$. In this sense, we may compare
the area scaling properties of quantum fluctuations to volume
scaling properties of statistical fluctuations in a canonical bulk
theory. From this point of view the area scaling of thermodynamic
quantities in restricted regions of space is due to quantum
entanglement. This is consistent with the observation that
entanglement entropy scales linearly with the boundary area
\cite{Bombelli,Srednicki}.

It has also been shown that one can relate spatial integrals over
n-point functions of a free field theory in the bulk, to those of
a free field theory on the boundary when the latter is half of
space \cite{holmin}. We wish to present further evidence that area
scaling of the variance of quantum fluctuations, and more general
correlation functions, should be ascribed to a boundary theory,
regardless of the details of the bulk theory or the geometry of
the boundary.

In what follows, we give several examples of energy fluctuations
in specific geometries. These examples highlight the differences
between area scaling and the usual notion of volume dependent
fluctuations obtained from canonical statistical systems. We
consider energy fluctuations since these are more intuitive to our
understanding, though in general many other operators also have
fluctuations which scale as the surface area. For example, Noether
charges associated to symmetries of the theory provide interesting
examples of bulk operators with surface fluctuations. A more
precise discussion of the general properties that characterize
bulk operators with area scaling fluctuations is presented in the
text.

To calculate energy fluctuations we consider the energy operator
of a certain spatial volume $V$ in d+1 dimensional Minkowski
space: $E^V=\int_V :\mathcal{H}(\vec{x}):d^dx$. The fluctuations
of this operator in its vacuum state is given by $(\Delta E^V)^2
\equiv \langle 0 | {(E^V)}^2 | 0 \rangle$, and may be evaluated
analytically for certain symmetric geometries, and numerically
otherwise. In any case, general considerations show that to
leading order, the fluctuations scale linearly with the boundary
area of $V$: $(\Delta E^V)^2 \propto S(V)$.

\section{The flower and the annulus}
To show the peculiarity of area scaling behavior (of energy
fluctuations) we consider geometries in which one would expect
that energy fluctuations decrease or remain the same if they are
volume dependent, and instead they increase. To keep things
simple, we work in 2+1 dimensions.

\subsection{The flower}
\begin{figure}
\begin{flushleft}
{\epsfig{file=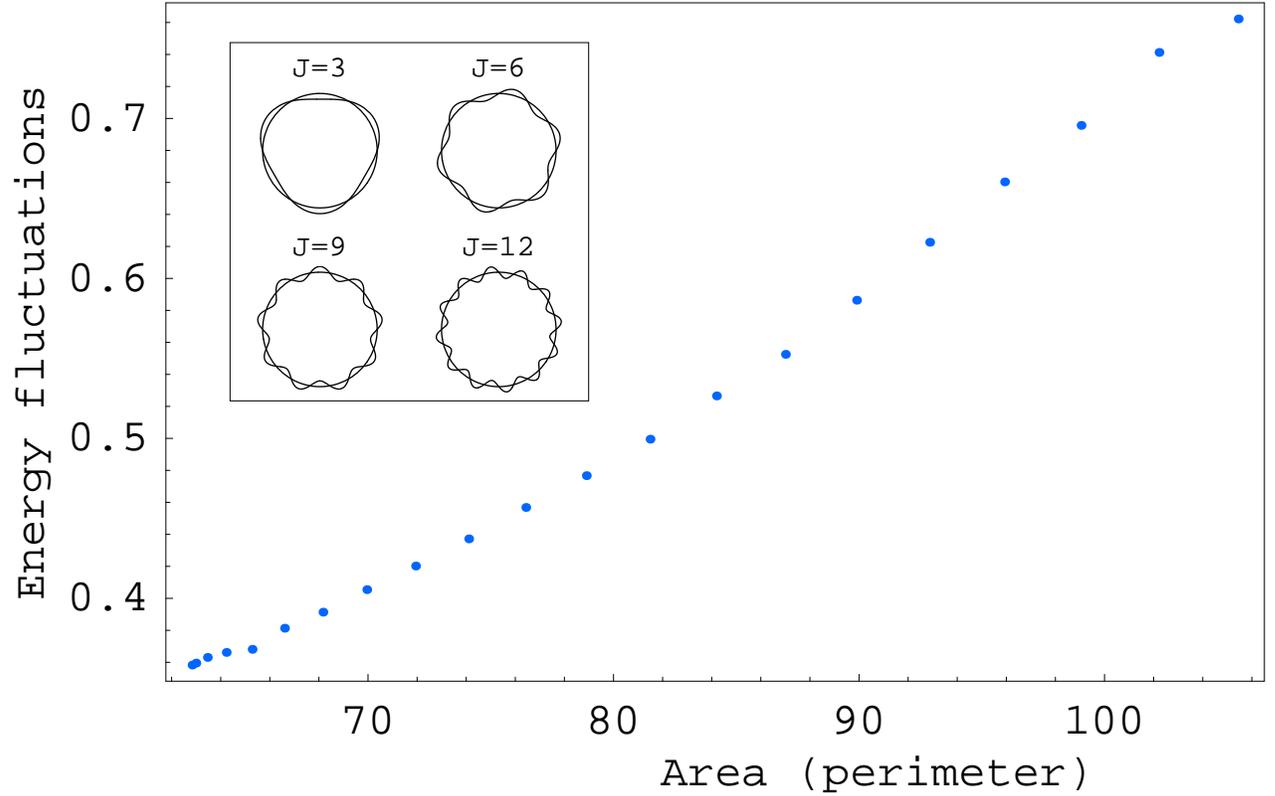,width=18.0cm}}
\end{flushleft}
\caption{Energy fluctuations in the flower geometry as a function
of surface area (perimeter). The horizontal axis represents the
area (perimeter in the 2-dimensional case) of the flower geometry.
The points on the plot represent the numerically evaluated variance
of energy fluctuations inside the flower as a function of the surface area
(perimeter). The boxed diagram at the top left, illustrates how
the surface area (perimeter) of the flower increases while the
volume (area) remains constant.} \label{F1}
\end{figure}

Our first example is the `flower' geometry \cite{Oaknin:2003dc}: a
shape whose boundary is given by $\vec{x}(\theta)=(R+\Delta R
\sin(J\theta))(\cos(\theta),\sin(\theta))$. For $J\neq 0$ the
surface area (length) of this shape increases with $J$ (the number
of `petals'), while the volume (area): $\pi(R^2+\frac{{\Delta
R}^2}{4})$ stays constant. If energy fluctuations were associated
to the bulk one would expect that they remain constant as $J$
increases. Instead, as the numerical calculation shows (See Figure
\ref{F1}), the variance of energy fluctuations increase as $J$ (and thus
the surface area) gets larger, suggesting that the fluctuations are
associated to the boundary.

\subsection{The annulus}
\begin{figure}
{\epsfig{file=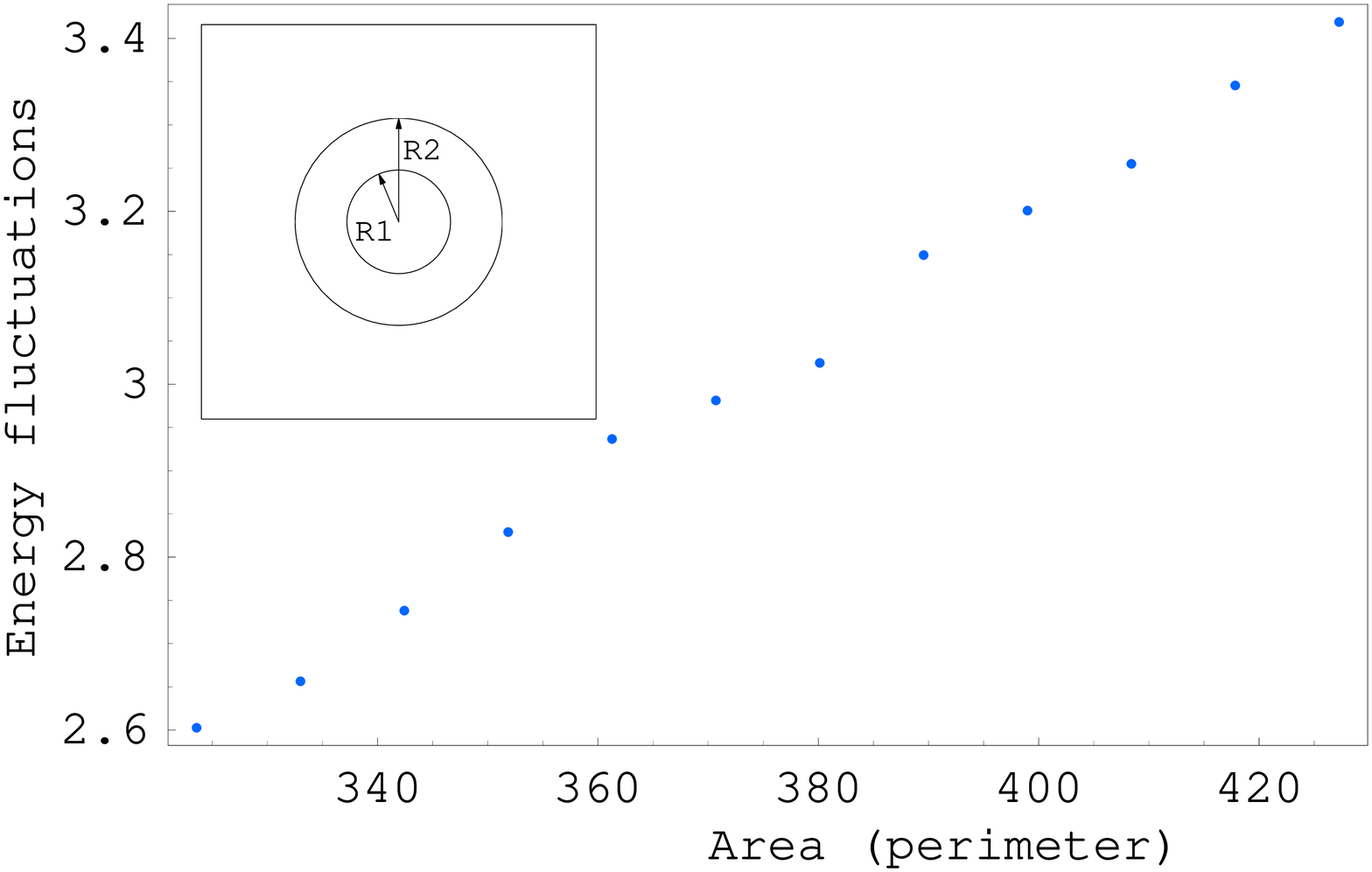,width=17.0cm}}\\
  \caption{Magnitude of the energy fluctuations as a function of the area (perimeter: $2\pi(R_1+R_2)$) of the annulus.
           The points on the graph show the numerically evaluated relation between the variance
           of the energy inside the annulus, and its surface area (perimeter). As the volume (area) of the annulus
           decreases, the energy fluctuations increase with the length of the perimeter.}
  \label{F2}
\end{figure}

A somewhat more radical situation is given by an annulus geometry.
We divide space into three regions. Region one is a circle of
radius $R_1$. Region two is an annulus of radii $R_1$ and $R_2$
($R_2>R_1$), which is concentric with the circle given earlier,
and region 3 is that part of space which encloses the annulus and
circle. As we increase the size of the inner radius $R_1$, the
volume of the annulus decreases, yet its surface area increases.
Therefore, when considering the energy fluctuations in the
annulus, we find, remarkably, that they increase as the annulus
becomes thinner. This is shown in figure \ref{F2}.

Another property that one may consider in an annulus
configuration, is the correlation between fluctuations in
different regions of space. Considering the volumes $V_1$ and
$V_2$ corresponding to the circle and annulus, we argue that the
covariance, $\text{cov}(E^{V_1},E^{V_2})$, of the energy in these
two regions is proportional to their mutual surface area (in this
case the circumference of the circle).

This can be stated as a more general argument. For any two
operators $O^{V_1}$ and $O^{V_2}$ defined as an integral over a
density $\int_{V_i} \mathcal{O} d^dx$,
$\text{cov}(O_{V_1},O_{V_2})$ scales linearly with the mutual
surface area, if the the two point function $F(r=|{\vec r}_1-{\vec
r}_2|)\equiv \langle 0|O({\vec r}_1) O({\vec r}_2) |0 \rangle =
\nabla^2 g(r)$ is derived from a function $g(r)$ which has short
range and is not too divergent ar $r=0$.

We give here an outline of the proof, whose details will be
presented elsewhere \cite{tbp}. We observe that the covariance of
these operators may be written as\footnote{For simplicity we
assume that the vacuum expectation value of $\mathcal{O}$ is
zero.}
\begin{align*}
    \text{cov}(O_{V_1},O_{V_2})
        &= \int_{V_1} \int_{V_2}
              \langle 0 | \mathcal{O}(\vec{r}_1)
                \mathcal{O}(\vec{r}_2) | 0 \rangle
           d^dr_1 \, d^dr_2\\
        &=  \int_{V_1} \int_{V_2}
               F(|\vec{r}_1-\vec{r}_2|)
           d^dr_1 \, d^dr_2\\
        &=  \int D_{V_1,V_2}(\xi) F(\xi) d\xi.
\end{align*}
All information on the geometry is contained in the function
\begin{equation}
\label{E:defofD}
    D_{V_1,V_2}(\xi)=\int_{V_1}
                \int_{V_2}
                    \delta(\xi-|\vec{r}_1-\vec{r}_2|)
                d^dr_1
                d^dr_2.
\end{equation}
Using $F=\nabla^2 g$, integrating by parts, and using the short
range behavior of $g$, one can show that the covariance of the two
operators scales, to leading order, as
$\tilde{D}(0)=\frac{d}{d\xi} \left( \frac{D(\xi)}{\xi^{d-1}}
\right) \Big |_{\xi=0}$~\cite{holmin}.

\begin{figure}
  \begin{center}
  \scalebox{0.35}{\includegraphics{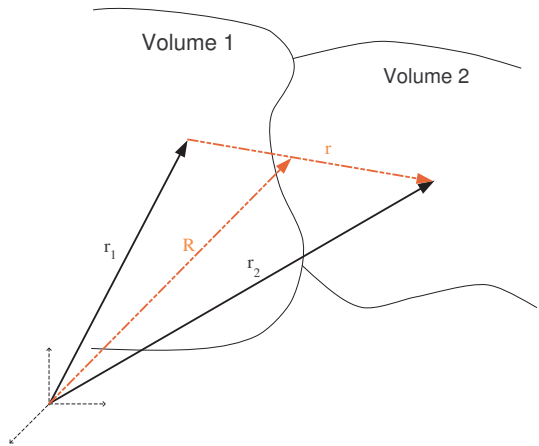}}\\
  \caption{Geometry of the $r_1$--$r_2$ coordinate system,
   and the $R$--$r$ coordinate system, when $V_1$ and $V_2$ have a mutual boundary
           but no mutual interior.}
  \label{F:geometry}
  \end{center}
\end{figure}

\begin{figure}
  \begin{center}
  \scalebox{0.35}{\includegraphics{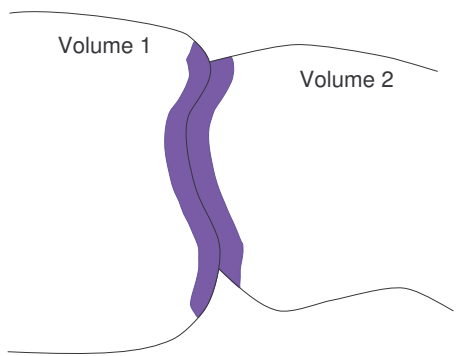}}\\
  \caption{The region of integration in the $R$--$r$ coordinate system
           when $V_1$ and $V_2$ have a mutual boundary but no mutual interior.}
  \label{F:geometry2}
  \end{center}
\end{figure}

We shall evaluate $\tilde{D}(\xi)$ for small $\xi$, where the
volumes $V_1$ and $V_2$ are just touching (from the outside) at a
boundary $B$. Let $\vec{r_i} \in V_i$, and consider the vectors
$\vec{R}=\frac{\vec{r}_1+\vec{r}_2}{2}$, and
$\vec{r}=\vec{r}_2-\vec{r}_1$. Geometrically $\vec{r}$ points from
$\vec{r}_1$ to $\vec{r}_2$, and $\vec{R}$ points to the center of
$\vec{r}$ (See figure \ref{F:geometry}). We may now do the
integration in (\ref{E:defofD}) in the $R$--$r$ coordinate system.
For a given value of $\xi$, we have $|\vec{r}|=\xi$, implying that
$\vec{r}_1$ and $\vec{r}_2$ are located such that the distance
between them is $\xi$. Since they are located at different ends of
the boundary, the vector $\vec{R}$ covers a volume of $S(B)\xi$
(where $S(B)$ is the surface area of $B$), centered at the
boundary between the two regions (See figure \ref{F:geometry2}).
Therefore, the integration over the $R$ coordinate in
(\ref{E:defofD}) will give $S(B)\xi$. The integral over the radial
part of the $r$ coordinate and the delta function will yield a
$\xi^{d-1}$ term, and the angular part, which is restricted as $R$
gets near the boundary, will contribute a geometric factor $G_S$.
Hence $\tilde{D}(0)=G_S S(B)$.

Therefore the fluctuations in the ring and the fluctuations in the
circle will be correlated such that their covariance is
proportional to their mutual surface area.

\section{Correlations of fluctuations}

We wish to generalize the above argument and show that the
covariance of any two operators $O_{V_1}$ and $O_{V_2}$ will be
proportional to the mutual surface area of the volumes $V_1$ and
$V_2$ (if, as stated earlier, the two point correlation function
$F(r)$ is derived from a short range function $g(r)$ which is not
too divergent ar $r=0$.) Again, we only outline the proof.

From the arguments of the previous section, it is enough to show
that $\tilde{D}(0)\propto S(B)$, with $B=B(V_{1}) \cap B(V_{2})$.
We consider three generic cases: first we look at a specific case
where the volume $V_1$ is contained completely (with no mutual
boundaries) in $V_2$. We switch to the $R$--$r$ coordinate system.
To leading order, the integration over the $\vec{R}$ coordinate
will be proportional to the volume, except for a region of
thickness $\xi$ from the boundary. To sub-leading order, there
will be contributions to the surface area term from within $V_1$
and external to $V_1$. The restriction of the angular part of
$\vec{r}$ inside $V_1$, and external to $V_1$ combine so that they
cancel the surface term introduced from the volume integration
over the $\vec{R}$ coordinate. So, in this case, there are no
surface area terms in $D(\xi)$, and therefore $\tilde{D}(0)=0$.

Next we look at regions $V_1$ and $V_2$ which have boundaries
overlapping only from within each other (that is $V_1 \subseteq
V_2$, and they have a mutual boundary). Consider $V_3$ which is
the complement of $V_2$, and define $O = O_{V_2}+O_{V_3}$. By the
comment made earlier, $\text{cov}(O,O_{V_1})=0$. This implies
$\text{cov}(O_{V_2},O_{V_1})+\text{cov}(O_{V_3},O_{V_1})=0$. Since
the boundaries of $V_3$ and $V_2$ are equal (by the definition of
a boundary, and assuming that the whole space has no boundary),
then $B \equiv B(V_3) \cap B(V_1)=B(V_2) \cap B(V_1)$. Since $V_3$
is external to $V_1$, we have from the earlier argument that
$\text{cov}(O_{V_3},O_{V_1})\propto G_S S(B)$, implying that
$\text{cov}(O_{V_2},O_{V_1})\propto -G_S S(B)$.

More complex geometries may now be handled by dividing $V_{1}$, (or $V_{2}$)
into subvolumes which satisfy the above conditions. This is treated in detail in \cite{tbp}.

Therefore, since $\tilde{D}(0) \propto S(B)$, we have that under
the restrictions mentioned, $\text{cov}(O^{V_{1}}O^{V_{2}})
\propto S(B)$. In what follows we give a numerical example of this
and discuss its implications.

\subsection{Displaced Boxes}

\begin{figure}
  {\epsfig{file=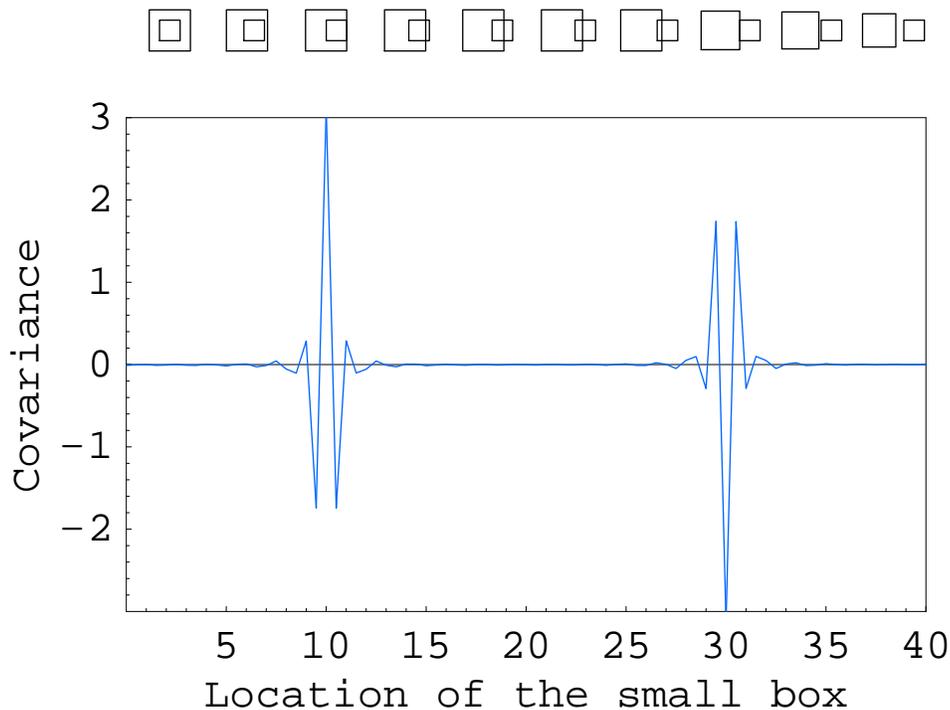,width=17.0cm}}\\
  \caption{Correlation between energies inside the boxes, as a function of the relative position of the boxes. Note
  the interference pattern as the distance between the surfaces becomes of the order of the U.V. cutoff.}
  \label{F3}
\end{figure}

Our final numerical example is that of relatively placed boxes. We
consider a large square box ${\cal B}$ of volume $L \times L$, and
inside it, a smaller square box ${\cal A}$ of volume $L/2 \times
L/2$. We then move the box ${\cal A}$ by an amount $\Delta x$
along one of the symmetry axes  of the box ${\cal B}$ as shown in
Figure~\ref{F3}.

In a bulk theory, we would expect that the statistical
fluctuations in the energy in the box ${\cal B}$, and those in the
box ${\cal A}$ be correlated only if the boxes have some mutual
volume, as there is no method by which, at a given time,
fluctuations in region ${\cal A}$ `know of' fluctuations in region
${\cal B}$. As the boxes are contained in each other, we expect to
see some correlation between fluctuations in both regions, and as
the boxes move farther apart the fluctuations will become less
correlated, until the regions are disjoint, at which point
measurements in region ${\cal A}$ and measurements in region
${\cal B}$ are not correlated.

However, as we showed above, when considering entanglement induced
fluctuations, the covariance will be zero, unless the boxes have a
mutual surface area. This is shown explicitly in the numerical
calculation (See Figure \ref{F3}).

The apparent interference pattern when the boundaries of the boxes
almost coincide is a result of U.V. effects---these attribute to
the boundaries a certain fuzziness which creates an interference
pattern when they almost coincide.

The correlations of the energy fluctuations in the boxes do not
behave as expected from correlations induced by a bulk theory:
constant when ${\cal A}$ is inside $\cal B$, and slowly decreasing
to zero as $\cal A$ exits $\cal B$. Instead the energy
fluctuations are uncorrelated, until the boxes have a mutual
surface. Therefore, the mechanism by which fluctuations in region
$\cal B$ and region $\cal A$ know of each other, must be dependent
on the boundary of the regions.

Also, the sign of the covariance changes when the surface is
`common' or `anti-common': apart from the energies being
correlated as the boundaries are in contact, we note that when the
boundaries are common from within we get that a positive
fluctuation in region ${\cal A}$ corresponds to a positive
fluctuation in region ${\cal B}$, whereas when the boundaries are
common from the outside, a positive fluctuation in ${\cal B}$
corresponds to a negative fluctuation in ${\cal A}$. This again,
is an indication that the fluctuations occur on the boundary.

\section{Summary}

We have given several examples of area scaling of energy
fluctuations in various geometries. The area fluctuations are
originally calculated in a bulk - type setting, yet they have
properties which are typical in boundary theories, suggesting that
perhaps  some corresponding boundary theory may give similar
results.

In the flower geometry, we have seen that the energy fluctuations
are very sensitive to the geometry of the boundary they are being
measured in. Had the fluctuations been those of a canonical
statistical ensemble, one would have expected that the
fluctuations of the energy be proportional to the volume of the
system, which in this case is constant.

In the annulus example the energy fluctuations in the annulus
increase with decreasing annulus volume (area) (but increasing
surface area (perimeter)). This, again, is in contradiction with
our usual understanding of bulk - type theories.

Finally, when considering the energy fluctuations in two different
regions of space, we have shown that these are correlated only if
the regions have a mutual boundary. Moreover, the sign of the
covariance changes as the boundaries are common or anti-common,
which, as explained earlier, is characteristic of a boundary
theory.

The surface dependence of fluctuations of extensive quantities
restricted to spatial sub-volumes is a characteristic feature of
glass-like statistical systems. In this paper we have discussed
some implications of this behaviour in the context of a free
quantum field theory in Minkowski space. Those conclusions could
be also of interest in the context of primordial cosmological
density fluctuations at the time of recombination, which share the
same glass-like character \cite{Oaknin:2003dc}.

\section{Acknowledgements}

Research supported in part by the Israel Science Foundation under
grant no. 174/00-2 and by the NSF under grant no.PHY-99-07949.
A.~Y. is partially supported by the Kreitman foundation. 
D.~O is partially supported by NSERC of Canada. R.~B.
thanks the KITP, UC at Santa Barbara, where this work has been
completed. We would like to thank B. Kol for discussions and B.
Reznik for clarifying some issues regarding entanglement and the
vacuum.

\begin{appendix}
\section{Numerical calculations}
For the numerical calculations we have considered a free scalar
field theory in 2+1 dimensional Minkowski space. We impose an I.R.
cutoff by defining all of space to have size $L \times L$ (with
periodic boundary conditions), and a U.V. momentum cutoff $\frac{2
\pi N}{L}$. The fields may be expanded in fourier modes:
\begin{align*}
    \Phi(\vec x) &= \sum_{n_1,n_2=-N}^{+N}
        \frac{1}{L} q_{(n_1,n_2)}
        exp\left(+2\pi \imath (\sum_{j=1}^{2} n_j x_j)/L \right)\\
    \Pi^{*}(\vec x) &= \sum_{n_1,n_2=-N}^{+N}
        \frac{1}{L} p^*_{(n_1,n_2)}
        exp\left(-2\pi \imath (\sum_{j=1}^{2} n_j x_j)/L \right).
\end{align*}
So that
\begin{equation*}
    E_V =\sum_{{\vec n},{\vec m}}
    \left( p^{*}_{\vec n} p_{\vec m} +
        \kappa^2_{{\vec n}\cdot{\vec m}} q^{*}_{\vec n} q_{\vec m}
    \right)
    A_V({\vec n}-{\vec m},L),
\end{equation*}
where
\begin{align*}
    \kappa^2_{{\vec n}\cdot{\vec m}} &= \left(\frac{2\pi}{L}\right)^2
        ({\vec n} \cdot {\vec m}) + M^2\\
\intertext{and}
    A_V({\vec k},L) &=
        \frac{1}{L^2} \int_V d^2{\vec x} \,
        exp \left(-2\pi \imath {\vec k}\cdot{\vec x}/L \right).
\end{align*}
Using the divergence theorem, one can simplify $A_V$
\begin{equation}
    A_V({\vec k},L) \equiv
    A_{\partial V}({\vec k},L) =
    \frac{1}{L^2} \int_{\partial V} \hspace{0.05in}
        \frac{iL}{2\pi} e^{-(2\pi \imath {\vec k}\cdot{\vec x}/L)}
        \frac{{\vec k} \cdot d{\vec {A}}}{|{\vec k}^2|}.
\end{equation}

The expectation value of $E_{V_1} E_{V_2}$ in the vacuum can be
expressed as
\[
    \text{var}(E_V)=
    \frac{1}{4} \sum_{{\vec n},{\vec m}}
        \left(
            \frac{\kappa^2_{{\vec n}\cdot{\vec m}}} {\sqrt{\kappa_{\vec n}} \sqrt{\kappa_{\vec m}}}
            - \sqrt{\kappa_{\vec n}} \sqrt{\kappa_{\vec m}}\right)^2
        |F_V({\vec n}-{\vec m},L)|^2,
\]
and that
\begin{multline*}
    \text{cov}(E_{V_1},E_{V_2})=
    \frac{1}{4} \sum_{{\vec n},{\vec m}}
        \left(\frac{\kappa^2_{{\vec n}\cdot{\vec m}}} {\sqrt{\kappa_{\vec n}} \sqrt{\kappa_{\vec m}}}
            -\sqrt{\kappa_{\vec n}} \sqrt{\kappa_{\vec m}}\right)^2  \times
%
    (F_{V1}({\vec n}-{\vec m},L))^*(F_{V2}({\vec n}-{\vec m},L)),
\end{multline*}
where we have defined
$\kappa^2_{\vec{n}}=\kappa^2_{\vec{n}\cdot\vec{n}}$. This
expression shows that the covariance is obtained as an
interference process between many different modes.

For the flower geometry we have used an I.R. cutoff of L=100
units, and a U.V. cutoff of N=25, yielding $\frac{L}{2\pi N}\sim
0.6$~units. As the distance between the petals decreases to the
order of the U.V. cutoff, one observes some slight deviation from
the area scaling law described above. A numerical calculation of
the flower geometry was carried out for free fields of masses
$M=0,0.5,1$~units$^{-1}$. The graph that we have presented in
Fig.~\ref{F1} corresponds to $\Delta R=1$~unit, and $R=10$~units.
We have also checked the results for free fields of masses $M=0,
0.5, 1, 2.5$~units$^{-1}$, radii R=10,20 and 40~units and $\Delta
R$ = 1,2, and 4~units. Apart from the deviations at large radii
which, as suggested earlier, are perhaps related to ``perimeter
corrections'', the results were similar.

The infrared cutoff for the annulus was fixed at L=100 units of
length and the ultraviolet cutoff by $\frac{L}{2\pi N} \simeq 1$
unit of length. The scalar field was taken to be massless. For the
plot given in Fig.~\ref{F2} we used an external radius of 40
units, and varied the inner radius from 11.5 units to 25.5 units.

For the relative boxes we have imposed an I.R. cutoff of L=100
units, and a U.V. cutoff using N=30, yielding $\frac{L}{2\pi
N}\sim 1.8$~units. Box ${\cal B}$ had a boundary located at
$(-20,20)\times(-20,20)$, and the smaller box's boundary was at
$(-10+\Delta x,10+\Delta x)\times(-10,10)$. Figure~\ref{F3} was
plotted for a field of mass $M=0.5$~units$^{-1}$. We get similar
results for a higher mass  field ($M=3.5$~units$^{-1}$).
\end{appendix}

\end{document}